\documentclass[
aps,%
12pt,%
final,%
notitlepage,%
oneside,%
onecolumn,%
nobibnotes,%
nofootinbib,%
superscriptaddress,%
showpacs,%
centertags]%
{revtex4}
\usepackage{amssymb,amsmath,epsfig}

\def\im{\mbox{Im\,}}
\def\re{\mbox{Re\,}}
\def\diag{\mbox{diag\,}}

\def\bbbc{{\Bbb C}}
\def\bbbr{{\Bbb R}}

\def\bbbs{{\Bbb S}}
\def\bbbz{{\Bbb Z}}
\def\openone{\leavevmode\hbox{\small1\kern-3.3pt\normalsize1}}
\newtheorem{remark}{{Remark}}%[section]

\def\bb {\begin {eqnarray}}
\def\ee {\end {eqnarray}}

\begin{document} %\selectlanguage{english}
\title
{$\bbbz_2$-reductions of spinor models in two dimensions}

\author{\firstname{V. S.}~\surname{Gerdjikov}}
\affiliation{Institute for Nuclear Research and Nuclear Energy,\\
Bulgarian Academy of Sciences\\ 1784 Sofia, Bulgaria}

\begin{abstract}
We propose new types of integrable spinor models, generalizing the
well known ones of: i) Nambu--Jona-Lasinio--Vaks--Larkin models, related to $SU(N)$;
ii) the Gross--Neveu models -- $SP(2N)$; and the iii) Zakharov--Mikhailov models
-- $SO(N)$. We propose a method for constructing their Lax representation and
outline the spectral properties of the Lax operators.
\end{abstract}

\maketitle
%{\bf \today }
\section{Introduction}

Spinor models play important role in contemporary theoretical physics.
The famous Nambu--Jona-Lasinio--Vaks--Larkin models \cite{NJL,VL}  and
Gross--Neveu models \cite{GN,NePa} have been proposed initially as models for describing the
strong interactions. Later, with the development of the inverse scattering method (ISM)
\cite{ZMNP,FaTa} it was proven that the two-dimensional versions of these models are
integrable \cite{zm1}. In the same paper Zakharov and Mikhailov propose a third class
of spinor models related to the orthogonal groups.

The aim of the present paper is to derive new types of
integrable spinor models by applying additional $\bbbz_2$-reductions
to their Lax representations. In doing this we will be using the reduction group \cite{mik}.
Next we describe the spectral properties of the Lax operators.

We start in Section II by some preliminaries concerning the spinor models and the reduction group
of Mikhailov \cite{mik}. In Section III we outline the spectral theory of the unreduced Lax operators.
In Section IV we derive the $\bbbz_2$-reduced spinor models. Section V is devoted to the spectral
properties of the $\bbbz_2$-reduced Lax operators. More specifically we treat 4 different cases
in each of which we specify the continuous spectrum and the symmetries of the discrete eigenvalues.
We end with brief conclusions.

\section{Preliminaries}

The integrability of the 2-dimensional versions of the  Nambu--Jona-Lasinio--Vaks--Larkin (NJLVL) and the
Gross--Neveu model  (GN) was discovered by Zakharov and Mikhailov in \cite{zm2}. The showed that
NJLVL models are related to   $su(N)$ algebras, while the Gross--Neveu models are related to the $sp(N)$.
In the same paper  an additional type of spinor models related to the algebras $so(N)$ was  discovered;
we will call them  Zakharov--Mikhailov (ZM) models.

Let us first outline the Lax representations of these models \cite{zm1}.

\begin{equation}\label{eq:U-Y}\begin{aligned}
\Psi_\xi & = U(\xi,\eta,\lambda) \Psi(\xi,\eta,\lambda), &\qquad \Psi_\eta & = U(\xi,\eta,\lambda) \Psi(\xi,\eta,\lambda),\\
U(\xi,\eta,\lambda) &= \frac{U_1(\xi,\eta)}{\lambda -a} , &\qquad
V(\xi,\eta,\lambda) &= \frac{V_1(\xi,\eta)}{\lambda -a} ,
\end{aligned}\end{equation}
where  $\eta =t+x$, $\xi=t-x$ and $a$ is a real number.

We also impose the $\bbbz_2$-reduction:
\begin{equation}\label{eq:red}\begin{split}
U^\dag (x,t,\lambda) &=-U(x,t,\lambda^*) , \qquad V^\dag (x,t,\lambda) =-V(x,t,\lambda^*).
\end{split}\end{equation}

The compatibility condition of the above linear problems reads:
\begin{equation}\label{eq:Lax}\begin{split}
 U_\eta - V_\xi + [U,V]=0,
\end{split}\end{equation}
which is equivalent to
\begin{equation}\label{eq:UV12}\begin{aligned}
U_{1,\eta} &+ \frac{1}{2a} [U_1, V_1(\xi,\eta)] =0,  \qquad
V_{1,\xi} &- \frac{1}{2a} [V_1, U_1(\xi,\eta)] =0.
\end{aligned}\end{equation}

From these equations, fixing up properly the gauge, (see \cite{zm2})  there follows that
\begin{equation}\label{eq:UVphi}\begin{aligned}
 U_1(\xi,\eta) & = -i\phi J_1^0 \phi^{-1},  \qquad
 V_1(\xi,\eta) & = i\psi I_1^0 \psi^{-1},
\end{aligned}\end{equation}
where $J_{1}^0$ and $I_{1}^0$ are properly chosen constant elements (choice of the gauge)
of the corresponding simple Lie algebra $\mathfrak{g}$.
In what follows we fix up $J_1^0 =-I_1^0 =J$ and choose $J$ for each of the above mentioned
models accordingly.
The matrix valued functions $\phi(\xi,\eta)$ and $\psi(\xi,\eta)$ take values in
the corresponding simple Lie group and are fundamental solutions of the following
ODE's:
\begin{equation}\label{eq:psifi2}\begin{aligned}
\psi_{\xi} &\equiv  - \frac{U_1(\xi,\eta)}{2a}  \psi(\xi,\eta) = \frac{i}{2a} \phi J \hat{\phi} \psi(\xi,\eta), \\
\phi_{\eta} &\equiv  \frac{V_1(\xi,\eta)}{2a}  \phi(\xi,\eta) = \frac{i}{2a} \psi J \hat{\psi} \phi(\xi,\eta).
\end{aligned}\end{equation}
Here and below by `hat' we will denote the inverse matrix, i.e. $\hat{\psi} \equiv (\psi)^{-1}$.

In this way we get three classes of spinor models. Below, following \cite{zm1} we  briefly outline
their derivation.

\begin{description}
  \item[i) Nambu-Jona-Lasinio-Vaks-Larkin models.] Here   we choose $\mathfrak{g}\simeq su(N)$. Then $\psi(\xi,\eta)$ and
  $\phi(\xi,\eta)$ are elements of the group $SU(N)$ and by definition $\hat{ \psi}(\xi,\eta) = \psi^\dag (\xi,\eta)$,
   $\hat{ \phi}(\xi,\eta) = \phi^\dag (\xi,\eta)$. Next we choose  $J=\diag(1,0,\dots,0)$ and as a result only the first
   columns $\phi^{(1)}$, $\psi^{(1)}$ and the first rows $\hat{ \phi}^{(1)}$, $\hat{ \psi}^{(1)}$ enter into the
   systems (\ref{eq:psifi2}). If we introduce the notations:
   \begin{equation}\label{eq:not1}\begin{split}
   \phi_\alpha (\xi,\eta) = \phi^{(1)}_{\alpha,1}, \qquad    \psi_\alpha (\xi,\eta) = \psi^{(1)}_{\alpha,1},
   \end{split}\end{equation}
    then the explicit form of the system is:
  \begin{equation}\label{eq:JLVS}\begin{aligned}
  \frac{\partial \phi_\alpha }{ \partial \eta } &=\frac{i}{2a} \psi_\alpha \sum_{\beta=1}^{N} \psi^*_{\beta}\phi_\beta , \\
  \frac{\partial \psi_\alpha }{ \partial \xi } &=\frac{i}{2a} \phi_\alpha \sum_{\beta=1}^{N} \phi^*_{\beta}\psi_\beta .
  \end{aligned}\end{equation}

  The functional of the action is:
  \begin{equation}\label{eq:Ljl}\begin{split}
  A_{\rm NJLVL} &= \int_{-\infty}^{\infty} dx\; dt\; \left( i \sum_{ \alpha=1}^{N} \left(\phi^*_{\alpha} \frac{\partial \phi_\alpha}{ \partial \eta }
  + \psi^*_{\alpha} \frac{\partial \psi_\alpha}{ \partial \xi } \right)
  - \frac{1}{2a} \left| \sum_{\alpha=1}^{N} ( \psi^*_{\alpha}\phi_\alpha  ) \right|^2  \right).
  \end{split}\end{equation}

  \item[ii) Gross-Nevew models.] Here we choose $\mathfrak{g}\simeq sp(2N,\bbbr)$; then $\psi(\xi,\eta)$ and
  $\phi(\xi,\eta)$ are elements of the group  $\mathfrak{G}\simeq SP(2N,\bbbr)$.
   Following \cite{zm1} we use the standard definition of symplectic group elements:
   \begin{equation}\label{eq:sym}\begin{split}
   \hat{ \psi}(\xi,\eta) = \mathfrak{J} \psi^T(\xi,\eta) \hat{ \mathfrak{J}}, \qquad
  \hat{ \phi}(\xi,\eta) = \mathfrak{J} \phi^T(\xi,\eta) \hat{ \mathfrak{J}}, \qquad \mathfrak{J} =
  \left(\begin{array}{cc} 0 & -\openone \\ \openone & 0    \end{array}\right).
   \end{split}\end{equation}
   Then the corresponding Lie algebraic elements acquire the following block-matrix structure:
   \begin{equation}\label{eq:spN}\begin{split}
   U_1(\xi, \eta) = \left(\begin{array}{cc} A & B \\ C & -A^T    \end{array}\right),
   \end{split}\end{equation}
   where $A,B,C$ are arbitrary real $N\times N$ matrices. Next we choose
   \begin{equation}\label{eq:J0}\begin{split}
    J =  \left(\begin{array}{cc} 0 & B_0 \\ 0 & 0      \end{array}\right), \qquad
    B_0 = \diag(1,0,\dots,0,0).
   \end{split}\end{equation}
    As a consequence again only the first columns $\phi^{(1)}$, $\psi^{(1)}$ and the first rows $\hat{ \phi}^{(1)}$, $\hat{ \psi}^{(1)}$ enter into the    systems (\ref{eq:psifi2}). If we introduce the $N$-component complex vectors:
   \begin{equation}\label{eq:not1'}\begin{split}
   \phi_\alpha (\xi,\eta) = \frac{1}{2} (\phi^{(1)}_{\alpha,1} +i \phi^{(1)}_{N+\alpha,1}), \qquad    \psi_\alpha (\xi,\eta)
   =\frac{1}{2} (\psi^{(1)}_{\alpha,1} +i \psi^{(1)}_{N+\alpha,1})
   \end{split}\end{equation}
    then the explicit form of the system is:
  \begin{equation}\label{eq:GN}\begin{aligned}
  \frac{\partial \phi_\alpha }{ \partial \eta } &= \frac{i}{a} \psi_\alpha \sum_{\beta=1}^{N}
  (\psi_{\beta}\phi^*_\beta -\psi^*_{\beta}\phi_\beta), \\
  \frac{\partial \psi_\alpha }{ \partial \xi } &=-\frac{i}{a} \phi_\alpha \sum_{\beta=1}^{N} (\phi_{\beta}\psi^*_\beta -
  \phi^*_{\beta}\psi_\beta).
  \end{aligned}\end{equation}

  The functional of the action is:
  \begin{equation}\label{eq:Lj2}\begin{split}
  A_{\rm GN} &= \int_{-\infty}^{\infty} dx\; dt\; \left( i \sum_{ \alpha=1}^{N} \left(\phi^*_{\alpha} \frac{\partial \phi_\alpha}{ \partial \eta }
    + \psi^*_{\alpha} \frac{\partial \psi_\alpha}{ \partial \xi }\right)
  - \frac{1}{2a} \left(\sum_{\alpha=1}^{N} ( \psi^*_{\alpha}\phi_\alpha -\phi^*_{\alpha} \psi_{\alpha}) \right)^2  \right).
  \end{split}\end{equation}

  \item[iii) Zakharov--Mikhailov models.] Now we choose $\mathfrak{g}\simeq so(N,\bbbr)$;
   then $\psi(\xi,\eta)$ and   $\phi(\xi,\eta)$ are elements of the group  $\mathfrak{G}\simeq SO(N,\bbbr)$.
   Following \cite{zm1} we use the standard definition of orthogonal group elements:
   \begin{equation}\label{eq:ort}\begin{split}
   \hat{ \psi}(\xi,\eta) =  \psi^T(\xi,\eta) , \qquad  \hat{ \phi}(\xi,\eta) =  \phi^T(\xi,\eta).
   \end{split}\end{equation}
    Now we choose
   \begin{equation}\label{eq:J0'}\begin{split}
    J =  E_{1,N} - E_{N,1},
   \end{split}\end{equation}
   where the $N\times N$ matrices $E_{kp}$ are defined by $(E_{kp})_{nm} =\delta_{kn} \delta_{pm}$.

    As a consequence now  the first and the last columns $\phi^{(1)}, \phi^{(N)}$, $\psi^{(1)}, \psi^{(N)}$
    and the first and the last rows $\hat{ \phi}^{(1)}, \hat{ \phi}^{(N)}$, $\hat{ \psi}^{(1)},
    \hat{ \psi}^{(N)}$ enter into the    systems (\ref{eq:psifi2}). If we introduce the $N$-component complex vectors:
   \begin{equation}\label{eq:not1''}\begin{split}
   \phi_\alpha (\xi,\eta) = \frac{1}{2} (\phi^{(1)}_{\alpha,1} +i \phi^{(N)}_{\alpha,N}), \qquad    \psi_\alpha (\xi,\eta)
   =\frac{1}{2} (\psi^{(1)}_{\alpha,1} +i \psi^{(N)}_{\alpha,N})
   \end{split}\end{equation}
  then the explicit form of the system becomes:
  \begin{equation}\label{eq:ZM}\begin{aligned}
  i\frac{\partial \psi_\alpha }{ \partial \xi } &=\frac{i}{a} \sum_{\beta=1}^{N} (\phi^*_{\alpha} \phi_{\beta} \psi_\beta -
  \phi_{\alpha} \phi^*_{\beta})\psi_\beta) , \\
  i\frac{\partial \phi_\alpha }{ \partial \eta } &=\frac{i}{a} \sum_{\beta=1}^{N} (\psi^*_{\alpha}  \psi_{\beta} \phi_\beta
 -\psi_{\alpha}  \psi^*_{\beta}) \phi_\beta ,
  \end{aligned}\end{equation}

  The functional of the action is:
  \begin{equation}\label{eq:Lj3}\begin{split}
  A_{\rm ZM} &= \int_{-\infty}^{\infty} dx\; dt\; \left( i \sum_{ \alpha=1}^{N} \left(\phi^*_{\alpha} \frac{\partial \phi_\alpha}{ \partial \eta }
    + \psi^*_{\alpha} \frac{\partial \psi_\alpha}{ \partial \xi }\right) \right. \\
  &- \left. \frac{1}{2a} \left(\sum_{\alpha,\beta =1}^{N} ( \phi^*_{\alpha}\phi_\beta -\phi^*_{\beta} \phi_{\alpha})
  ( \psi^*_{\alpha}\psi_\beta -\psi^*_{\beta} \psi_{\alpha}) \right)  \right).
  \end{split}\end{equation}

\end{description}
For more details of deriving the models see \cite{zm1}.

\section{ Spectral properties of the Lax operator}

Here we briefly outline the construction of the fundamental analytic solutions
of the Lax operator $L$.

First we fix up the class of potentials $U_1(\xi,\eta)$ and $V_1(\xi,\eta)$ by assuming
that $U_1(\xi,\eta)+iJ$ and $V_1(\xi,\eta)-iJ$ are Schwartz-type functions of
$\xi$ and $\eta$. We also assume that $J\in \mathfrak{h}$ is a real element of the Cartan
subalgebra of $\mathfrak{g}$.

\begin{remark}\label{rem:1}
These conditions are compatible with two of the classes of spinor models
listed above. These are, first  of all the NJLVL models for which $J$,
up to a trivial term $1/N \openone$, belongs to the Cartan subalgebra of $su(n)$.
For the ZM models $J$ belongs to the Cartan subalgebra, which in that case
consists of off-diagonal matrices. However, there is a simple similarity transformation
which takes $J$ of eq. (\ref{eq:J0'}) into $J= i {\rm \diag} (1,0,\dots,0 ,-1)$.

For the GN models the choice of $J$ is nilpotent. The spectral problem for such
Lax operators is singular and will not be discussed here.

\end{remark}

In what follows we will consider the spectral problem for Lax operators of the type:
\begin{equation}\label{eq:lax}\begin{split}
L\Psi (\xi,\eta,\lambda) \equiv \frac{\partial \Psi}{ \partial \xi } + i\frac{\phi J \hat{\phi}}{\lambda -a} \Psi(\xi,
\eta,\lambda)=0,
\end{split}\end{equation}
where $J\in \mathfrak{h}$ and $\phi (\xi,\eta) \in \mathfrak{G}$ and $\lim_{\xi\to\pm\infty} \phi (\xi,\eta) =\openone $.
The Jost solutions of $L$ are defined by:
\begin{equation}\label{eq:Jo}\begin{split}
\lim_{\xi\to\infty} \Psi_+(\xi,\eta,\lambda) \hat{\mathcal{E}}(\xi,\lambda) &=\openone, \qquad
\lim_{\xi\to -\infty} \Psi_-(\xi,\eta,\lambda) \hat{\mathcal{E}}(\xi,\lambda) =\openone,
\end{split}\end{equation}
where
\begin{equation}\label{eq:E}\begin{split}
\mathcal{E}(\xi,\lambda) &=\exp \left( -i \frac{ J\xi}{\lambda -a} \right).
\end{split}\end{equation}
The scattering matrix is introduced by:
\begin{equation}\label{eq:T}\begin{split}
T(\lambda,\eta) = \hat{\Psi}_+(\xi,\eta,\lambda) \Psi_-(\xi,\eta,\lambda)
\end{split}\end{equation}

\begin{figure}
  \includegraphics[width=6cm]{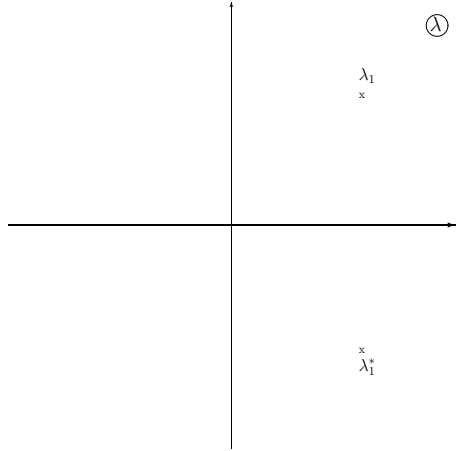}
  \caption{The continuous and the discrete spectrum of the operators $L$.  }\label{fig:0}
\end{figure}

The continuous spectrum of $L$ is located on a line of the complex $\lambda$-plane on which
$\mathcal{E}(\xi,\lambda)$ oscillates. In our case the continuous spectrum of $L$
fills up the real axis on the complex $\lambda$-plane. The discrete eigenvalues $\lambda_k^\pm \in \bbbc_\pm$
come in pairs, which due to the reduction (\ref{eq:red})  are mutually conjugate
$\lambda_k^+ =(\lambda_k^-)^*$, see fig. \ref{fig:0}.

The next step is to construct the fundamental analytic solutions (FAS) of $L$.
Their construction is done analogously to the case of the generalized Zakharov-Shabat system,
see \cite{GKV*09}. To this end we need the generalized Gauss decomposition of $T(\lambda,\eta)$
compatible with $J$:
\begin{equation}\label{eq:gau}\begin{split}
T(\lambda,\eta) = T_J^-D_J^+ \hat{S}_J^+, \qquad T(\lambda,\eta) = T_J^+D_J^- \hat{S}_J^-,
\end{split}\end{equation}
If $J=\diag (1,0,\dots,0)$ then
\begin{equation}\label{eq:TSJ0}\begin{aligned}
S_J^+(\eta,\lambda) & = \left(\begin{array}{cc}1 & \vec{s}\;^{+,T} \\ 0 & \openone  \end{array}\right), &\qquad
S_J^-(\eta,\lambda) & = \left(\begin{array}{cc}1 & 0\\ \vec{s}^-  & \openone  \end{array}\right), \\
T_J^+(\eta,\lambda) & = \left(\begin{array}{cc}1 & \vec{\tau}\;^{+,T} \\ 0 & \openone  \end{array}\right), &\qquad
S_J^-(\eta,\lambda) & = \left(\begin{array}{cc}1 & 0\\ \vec{\tau}^-  & \openone  \end{array}\right), \\
D_J^+(\lambda) & = \left(\begin{array}{cc}d_1^+  & \\ 0 & {\bf d}_2^+   \end{array}\right), &\qquad
D_J^-(\lambda) & = \left(\begin{array}{cc}d_1^-  & \\ 0 & {\bf d}_2^-    \end{array}\right).
\end{aligned}\end{equation}

For $J=\diag (1,0,\dots,0,-1)$ and $\mathfrak{g}\simeq so(N)$ we have:
\begin{equation}\label{eq:TSJ1}\begin{aligned}
S_J^+(\eta,\lambda) & = \left(\begin{array}{ccc}1 & \vec{s}\;^{+,T} & s^{\prime,+}\\ 0 & \openone
& s_0 \vec{s}\;^+ \\ 0 & 0 &1 \end{array}\right), &\qquad
S_J^-(\eta,\lambda) & = \left(\begin{array}{ccc}1 & 0 &0 \\ \vec{s}^-  & \openone &0 \\
s^{\prime,-} & \vec{s}\;^{-,T}s_0 & 1 \end{array}\right), \\
T_J^+(\eta,\lambda) & = \left(\begin{array}{ccc}1 & \vec{\tau}\;^{+,T} & \tau^{\prime,+} \\ 0 & \openone
& s_0 \vec{\tau}\;^+ \\ 0 & 0 1 \end{array}\right), &\qquad
S_J^-(\eta,\lambda) & = \left(\begin{array}{ccc}1 & 0 &0 \\ \vec{\tau}^-  & \openone &0 \\
\tau^{\prime,-} & \vec{\tau}\;^{-,T} s_0 & 1  \end{array}\right), \\
D_J^+(\lambda) & = \left(\begin{array}{ccc}d_1^+  & 0 & 0 \\ 0 & {\bf d}_2^+  &0 \\
0 & 0 & 1/d_1^+ \end{array}\right), &\qquad
D_J^-(\lambda) & = \left(\begin{array}{ccc}d_1^-  & 0 & 0 \\ 0 &  {\bf d}_2^-  &0 \\
0 & 0 & 1/d_1^- \end{array}\right),
\end{aligned}\end{equation}
where $s_0 = \sum_{j=1}^{N} E_{j,N+1-j}$.

Then the FAS analytic for $\lambda\in \bbbc_\pm$ are related to Jost solutions by:
\begin{equation}\label{eq:fas}\begin{split}
\chi^\pm(\xi,\eta,\lambda) = \Psi_-(\xi,\eta,\lambda)S_J^\pm (\eta,\lambda)= \Psi_+(\xi,\eta,\lambda)T_J^\pm (\eta,\lambda) D_J^\pm (\lambda).
\end{split}\end{equation}

The FAS  (\ref{eq:fas}) satisfy a Riemann-Hilbert problem (RHP) with canonical normalization at $\lambda \to a$:
\begin{equation}\label{eq:rhp}\begin{split}
\chi^+(\xi,\eta,\lambda) &= \chi^-(\xi,\eta,\lambda) G_J(\lambda,\eta), \qquad G_J(\lambda,\eta)= \hat{S}_J^-(\lambda,\eta)
S_J^+(\lambda,\eta), \\
\lim_{\lambda\to a} \chi^+(\xi,\eta,\lambda) &=\openone.
\end{split}\end{equation}
The canonical normalization of the RHP means that the FAS allow asymptotic expansions over the powers of $\lambda -a$:
\begin{equation}\label{eq:asXi}\begin{split}
\chi^\pm (\xi,\eta,\lambda) = \openone + \sum_{ s=1}^{\infty} X_s^\pm (\xi,\eta) (\lambda -a)^{-s} .
\end{split}\end{equation}
Therefore if we are given a solution of the RHP $\chi^+(\xi,\eta,\lambda)$ then the corresponding
potential of $L$ can be recovered from
\begin{equation}\label{eq:XiU}\begin{split}
U_1(\xi,\eta) \equiv igJg^{-1}(\xi,\eta) = i \frac{\partial X_1^\pm}{ \partial \xi }.
\end{split}\end{equation}

We finish this Section with the obvious remark, that the RHP formulation allows one
to derive the $N$-soliton solutions of the corresponding model via the Zakharov-Shabat
dressing procedure \cite{zm1}.

\section{ $\bbbz_2$-Reductions  of the spinor models}

Here we combine the construction of spinor models in two dimensions \cite{zm2}
with the idea of the reduction group \cite{mik}. Thus we intend to construct new types of
spinor models generalizing the ones in Section II.

Start with the Lax representation:
\begin{equation}\label{eq:U-YR}\begin{aligned}
\Psi_\xi & = U_{\rm R}(\xi,\eta,\lambda) \Psi(\xi,\eta,\lambda), &\qquad
\Psi_\eta & = V_{\rm R}(\xi,\eta,\lambda) \Psi(\xi,\eta,\lambda),\\
U_{\rm R}(\xi,\eta,\lambda) &= \frac{U_1(\xi,\eta)}{\lambda -a} + \frac{C U_1(\xi,\eta) C^{-1}}{\epsilon\lambda^{-1} -a} , &\qquad
V_{\rm R}(\xi,\eta,\lambda) &= \frac{V_1(\xi,\eta)}{\lambda +a} + \frac{CV_1(\xi,\eta) C^{-1}}{\epsilon\lambda^{-1} +a} ,
\end{aligned}\end{equation}
where $\epsilon =\pm 1$, $a\neq 1$ is a real number and $C$ is an involutive automorphism of $\mathfrak{g}$.
Obviously this Lax representation along with the typical reduction (\ref{eq:red}) satisfy also:
\begin{equation}\label{eq:UVR}\begin{aligned}
U_{\rm R}(\xi,\eta,\lambda) &= CU_{\rm R}(\xi,\eta,\epsilon\lambda^{-1}) C^{-1}, &\qquad
V_{\rm R}(\xi,\eta,\lambda) &= CV_{\rm R}(\xi,\eta,\epsilon\lambda^{-1}) C^{-1},
\end{aligned}\end{equation}
which is automatically compatible with the Lax representation \cite{mik}.

The new Lax representation is:
\begin{equation}\label{eq:LaxR}\begin{split}
\frac{\partial U_{\rm R}}{ \partial \eta}  - \frac{\partial V_{\rm R}}{ \partial \xi} + [U_{\rm R},V_{\rm R}]=0,
\end{split}\end{equation}
which is equivalent to
\begin{equation}\label{eq:UV12R}\begin{aligned}
U_{1,\eta} &+ [U_1, V_{\rm R}(\xi,\eta,a)] =0, &\qquad V_{1,\xi} &+ [V_1, U_{\rm R}(\xi,\eta,-a)] =0.
\end{aligned}\end{equation}

Next we apply the same way of deriving the models as in Section II; obviously, due to the additional
terms in $U_{\rm R}$ and $V_{\rm R}$ we get additional terms in the models.
In what follows we also list some typical choices for the automorphism $C$.
Skipping the details we get:

\begin{description}
  \item[i) $\bbbz_2$-NJLVL models.]
Here $\mathfrak{G}\simeq SU(N)$ and the system takes the form:
\begin{equation}\label{eq:phiR}\begin{aligned}
i\frac{\partial \vec{\phi}}{ \partial \eta} &+\frac{1}{2a} \vec{\psi}  ({\vec{\psi}\;}^\dag \vec{\phi})
+\frac{1}{\epsilon a^{-1}+a} C \vec{\psi} ({\vec{\psi}\;}^\dag \hat{C} \vec{\phi})(\xi,\eta) =0, \\
i\frac{\partial \vec{\psi}}{ \partial \xi}  & + \frac{ 1}{2a} \vec{\phi}  ({\vec{\phi}\;}^\dag \vec{\psi} )
+\frac{ 1}{\epsilon a^{-1}+a} C\vec{\phi} ({\vec{\phi}\;}^\dag \hat{C} \vec{\psi})(\xi,\eta) =0.
\end{aligned}\end{equation}
where $\vec{\psi} = (\psi_{\alpha,1} , \dots ,\psi_{\alpha,N})^T$ and $\vec{\phi} = (\phi_{\alpha,1} , \dots ,\phi_{\alpha,N})^T$.

For the automorphism $C$ of the $SU(N)$ group we may have
\begin{equation}\label{eq:CN-su}\begin{split}
\mbox{a)} \qquad C_N = \diag (\epsilon_1, \epsilon_2, \dots ,\epsilon_N), \qquad \epsilon_j=\pm 1, \qquad
\mbox{b)} \qquad C'_N = \left(\begin{array}{cc} 1 & 0  \\ 0 & C_{N-1}   \end{array}\right).
\end{split}\end{equation}
where $C_{N-1}$ belongs to the Weyl group of $SU(N-1)$ and is such that $C_{N-1}^2 =\openone$.
These two special choices of $C$ are such that $\lim_{\xi\to\pm\infty} U_R(\xi,\eta) =
\lim_{\xi\to\pm\infty} CU_R(\xi,\eta) \hat{C}$.

\item[ii)  $\bbbz_2$-GN models. ] Here $\mathfrak{G}\simeq SP(2N,\bbbr)$ and the form of the reduced system
depends on the choice of the automorphism $C$. Two  typical choices of $C$ are given by:
\begin{equation}\label{eq:CN-so}\begin{aligned}
\mbox{a)} \qquad C  &= \left(\begin{array}{cc} C_1 & 0 \\ 0 & C_1  \end{array}\right), &\qquad
\mbox{b)} \qquad C'  &= \left(\begin{array}{cc} 0 & C_2 \\  C_{2} & 0   \end{array}\right),
\end{aligned}\end{equation}
where $C_1^2 =C_2^2 =\openone$. In this way we obtain two different systems of GN-type.
Using  the $N$-component vectors $\vec{\psi}$ and $\vec{\phi}$ we can write them down in compact form:
  \begin{equation}\label{eq:GNz2a}\begin{aligned}
  \frac{\partial \vec{\phi}  }{ \partial \eta } &= -\frac{i}{a} \vec{\psi}
  \left( ( \vec{\psi}\;^\dag ,\vec{\phi})  - (\vec{\phi}\;^\dag ,\vec{\psi} ) \right)
  - \frac{2i}{a +\epsilon a^{-1}} C_1\vec{\psi}
  \left( ( \vec{\psi}\;^\dag C_1\vec{\phi})  - (\vec{\phi}\;^\dag C_1 \vec{\psi} ) \right)  , \\
  \frac{\partial \vec{\psi}  }{ \partial \xi } &= \frac{i}{a} \vec{\phi}
  \left( ( \vec{\psi}\;^\dag ,\vec{\phi})  - (\vec{\phi}\;^\dag ,\vec{\psi} ) \right)
  +\frac{2i}{a +\epsilon a^{-1}} C_1\vec{\phi}
  \left( ( \vec{\psi}\;^\dag C_1\vec{\phi})  - (\vec{\phi}\;^\dag C_1 \vec{\psi} ) \right).
  \end{aligned}\end{equation}
  The corresponding  action can be written   as follows:
  \begin{multline}\label{eq:AGNa}
  A_{\bbbz_2,\rm GNa} = \int_{-\infty}^{\infty} dx\; dt\; \left( i  \left(\vec{\phi}\;^\dag  \frac{\partial \vec{\phi}}{ \partial \eta }
    + \vec{\psi}\;^\dag \frac{\partial \vec{\psi} }{ \partial \xi }\right)
 - \frac{1}{2a} \left(( \vec{\psi}\;^\dag, \vec{\phi}) -(\vec{\phi}\;^\dag ,\vec{\psi} ) \right)^2 \right. \\
      \left. -\frac{1}{\epsilon a^{-1} +a} \left(( \vec{\psi}\;^\dag C_1 \vec{\phi}) -(\vec{\phi}\;^\dag C_1\vec{\psi} ) \right)^2    \right).
  \end{multline}

The second $\bbbz_2$-reduced GN-system is:
  \begin{equation}\label{eq:GNz2b}\begin{aligned}
  \frac{\partial \vec{\phi}  }{ \partial \eta } &= -\frac{i}{a} \vec{\psi}
  \left( ( \vec{\psi}\;^\dag ,\vec{\phi})  - (\vec{\phi}\;^\dag ,\vec{\psi} ) \right)
  + \frac{2i}{a +\epsilon a^{-1}} C_2\vec{\psi}\;^*
  \left( ( \vec{\psi}^T C_2\vec{\phi})  + (\vec{\psi}\;^\dag C_2 \vec{\phi}\;^* ) \right)  , \\
  \frac{\partial \vec{\psi}  }{ \partial \xi } &= \frac{i}{a} \vec{\phi}
  \left( ( \vec{\psi}\;^\dag ,\vec{\phi})  - (\vec{\phi}\;^\dag ,\vec{\psi} ) \right)
  +\frac{2i}{a +\epsilon a^{-1}} C_2\vec{\phi}\;^*
  \left( ( \vec{\phi}^T C_2\vec{\psi})  + (\vec{\phi}\;^\dag C_2 \vec{\psi}\;^* ) \right).
  \end{aligned}\end{equation}
These equations can be obtained from the action:
  \begin{multline}\label{eq:AGNb}
  A_{\bbbz_2,\rm GNb} = \int_{-\infty}^{\infty} dx\; dt\; \left( i  \left(\vec{\phi}\;^\dag  \frac{\partial \vec{\phi}}{ \partial \eta }
    + \vec{\psi}\;^\dag \frac{\partial \vec{\psi} }{ \partial \xi }\right)
 - \frac{1}{2a} \left(( \vec{\psi}\;^\dag, \vec{\phi}) -(\vec{\phi}\;^\dag ,\vec{\psi} ) \right)^2 \right. \\
      \left. -\frac{1}{\epsilon a^{-1} +a} \left(( \vec{\phi}\;^\dag C_2 \vec{\psi}\;^*) +(\vec{\phi}^T C_2\vec{\psi} ) \right)^2    \right).
  \end{multline}

  \item[iii) $\bbbz_2$-ZM models.] Here $\mathfrak{G}\simeq SO(N,\bbbr)$. Again we used
  $N$-component vectors to cast the $\bbbz_2$-reduced ZM systems in the form:
  \begin{equation}\label{eq:ZMa}\begin{aligned}
  \frac{\partial \vec{\psi}  }{ \partial \xi } &= \frac{i}{a} \left( \vec{\phi}\;^*
  ( \vec{\phi}^T ,\vec{\psi})  - \vec{\phi} (\vec{\phi}\;^\dag ,\vec{\psi} ) \right)
  + \frac{2i}{a +\epsilon a^{-1}} C \left( \vec{\phi}\;^*
   ( \vec{\phi}^T C\vec{\psi}) - \vec{\phi}(\vec{\phi}\;^\dag C \vec{\psi} ) \right)  , \\
  \frac{\partial \vec{\phi}  }{ \partial \eta } &= \frac{i}{a} \left( \vec{\psi}\;^*
   ( \vec{\psi}^T ,\vec{\phi})  - \vec{\psi} (\vec{\psi}\;^\dag ,\vec{\phi} ) \right)
  +\frac{2i}{a +\epsilon a^{-1}} C \left( \vec{\psi}\;^*
   ( \vec{\psi}^T \hat{ C} \vec{\phi})  - \vec{\psi} (\vec{\psi}\;^\dag \hat{ C}\vec{\phi} ) \right),
  \end{aligned}\end{equation}
where  the involutive automorphism $C$ can be chosen as one of the type:
\begin{equation}\label{eq:ZMC}\begin{aligned}
\mbox{a)} \qquad C &= \diag (\epsilon_1, \epsilon_2,  \dots , \epsilon_2,\epsilon_1   ), \qquad \epsilon_j=\pm 1, \qquad
\mbox{b)} \qquad C'  &= \left(\begin{array}{cc} 1 & 0 \\ 0 & C_3   \end{array}\right),
\end{aligned}\end{equation}
with $C_3^2=\openone$. For these choices of $C$ we have $\lim_{\xi\to\pm\infty} U_R(\xi,\eta) =
\lim_{\xi\to\pm\infty} CU_R(\xi,\eta) \hat{C}$.

The action for the reduced ZM models is provided by:
  \begin{multline}\label{eq:AGNbA}
  A_{\bbbz_2,\rm ZM} = \int_{-\infty}^{\infty} dx\; dt\; \left( i  \left(\vec{\phi}\;^\dag  \frac{\partial \vec{\phi}}{ \partial \eta }
    + \vec{\psi}\;^\dag \frac{\partial \vec{\psi} }{ \partial \xi }\right)
  +\frac{1}{a} \left(( \vec{\psi}\;^\dag, \vec{\phi}\;^*) (\vec{\phi}^T ,\vec{\psi}) -
  ( \vec{\phi}\;^\dag, \vec{\psi}) (\vec{\psi}\;^\dag ,\vec{\phi}) \right) \right. \\
      \left. +\frac{2}{\epsilon a^{-1} +a}
      \left(( \vec{\psi}\;^\dag C \vec{\phi}\;^*) (\vec{\phi}^T C\vec{\psi}) -
  ( \vec{\phi}\;^\dag C \vec{\psi}) (\vec{\psi}\;^\dag C\vec{\phi})  \right)    \right).
  \end{multline}

\end{description}

\section{ Spectral properties of the reduced Lax operators}

Here we briefly outline the construction of the fundamental analytic solutions
of the Lax operator $L_{\rm R}$. First we introduce the Jost solutions:
\begin{equation}\label{eq:E-R}\begin{split}
\lim_{\xi\to\infty} \Psi_{\rm R, +}(x,t,\lambda) \mathcal{E}_{\rm R}^{-1}(x,t,\lambda) &=\openone, \qquad
\lim_{\xi\to -\infty} \Psi_{\rm R, -}(x,t,\lambda) \mathcal{E}_{\rm R}^{-1}(x,t,\lambda) =\openone,\\
\mathcal{E}_{\rm R}(x,t,\lambda) &=\exp \left( -i \frac{ J\xi}{\lambda -a}  -i \frac{ CJC^{-1}\xi}{\epsilon\lambda^{-1} -a}
 \right),
\end{split}\end{equation}

The scattering matrix is defined by:
\begin{equation}\label{eq:TR}\begin{split}
T_{\rm R}(\lambda,\eta) = \hat{\Psi}_{\rm R,+}(x,t,\lambda) \Psi_{\rm R -}(x,t,\lambda)
\end{split}\end{equation}

Again we will need the generalized Gauss decomposition compatible with $J$:
\begin{equation}\label{eq:gauR}\begin{split}
T_{\rm R}(\lambda,t) = T_{J,\rm R}^-D_{J,\rm R}^+ \hat{S}_{J,\rm R}^+, \qquad T_{\rm R}(\lambda,t) = T_{J,\rm R}^+D_{J,\rm R}^-
\hat{S}_{J,\rm R}^-,
\end{split}\end{equation}
Their block-matrix form is the same like in eq. (\ref{eq:TSJ0}) or  (\ref{eq:TSJ1}); the only difference is that they
should satisfy the additional symmetry condition with respect to the second involution of $L_{\rm R}$.

The continuous spectrum of $L_{\rm R}$ (\ref{eq:U-YR}) fills up the curves on the complex $\lambda$-plane on which
\begin{equation}\label{eq:cs}\begin{split}
\re \left( -i \frac{ J\xi}{\lambda -a}  -i \frac{ CJC^{-1}\xi}{\epsilon\lambda^{-1} -a}  \right) = \im
\left(  \frac{ J\xi}{\lambda -a}  + \frac{ CJC^{-1}\xi}{\epsilon\lambda^{-1} -a}  \right) =0.
\end{split}\end{equation}
Below we  consider four different cases depending on the choice of $\epsilon$ and $C$.
 For convenience we denote $\lambda =\lambda_0 +i\lambda_1$ where $\lambda_0$
and $\lambda_1$ are real.

\begin{figure}
  \includegraphics[width=6cm]{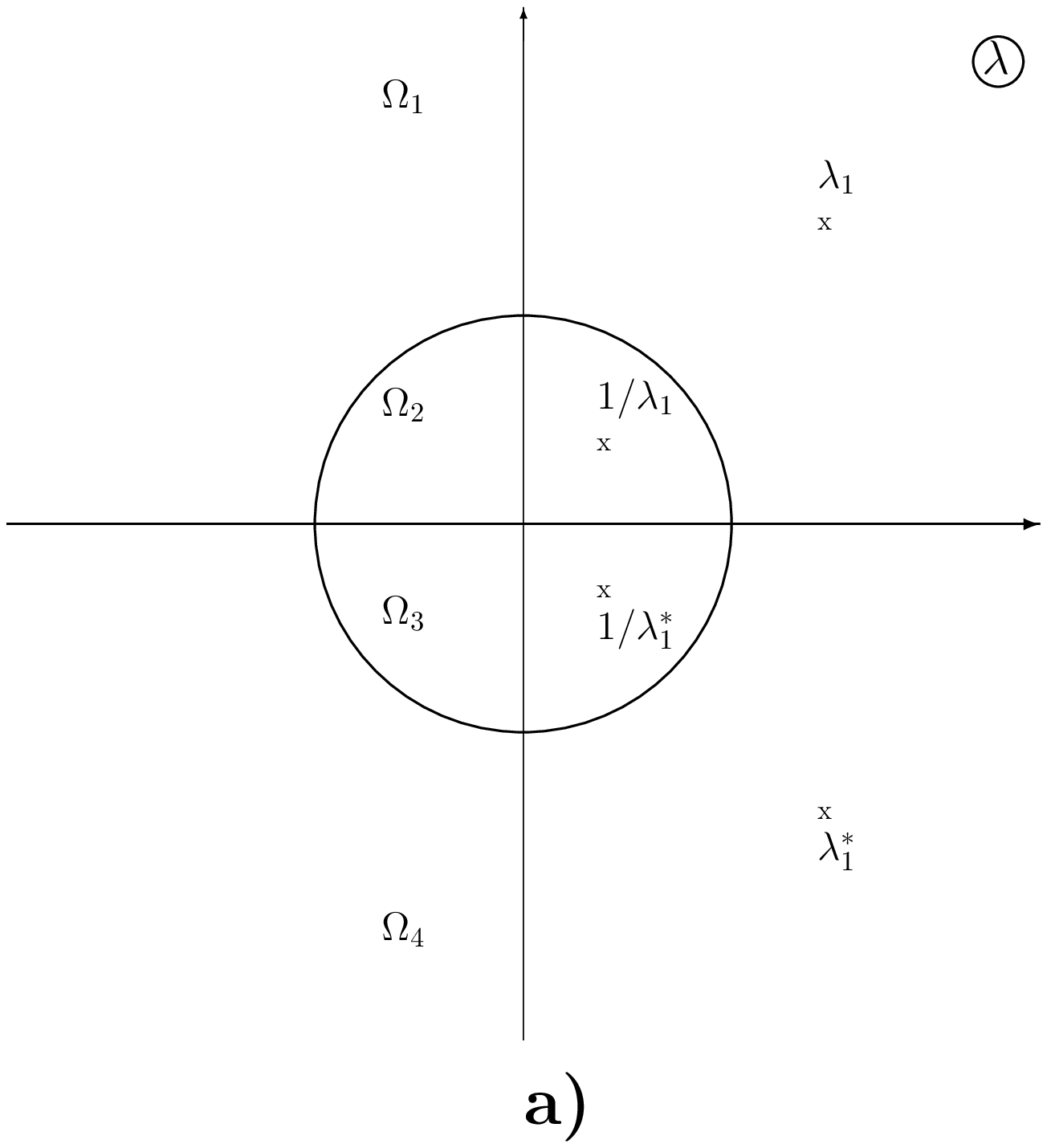}\qquad\includegraphics[width=6cm]{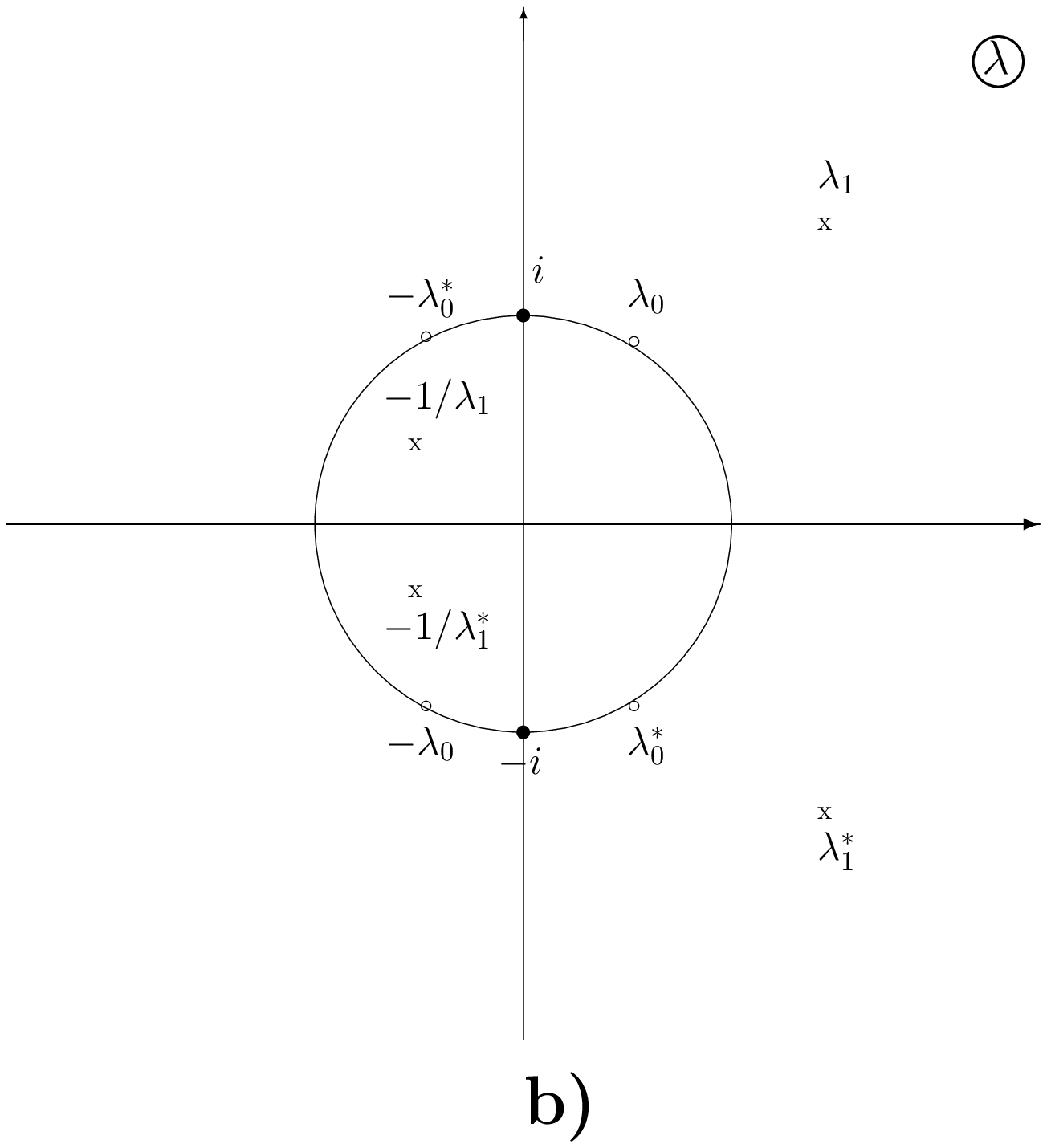}\\
  \includegraphics[width=6cm]{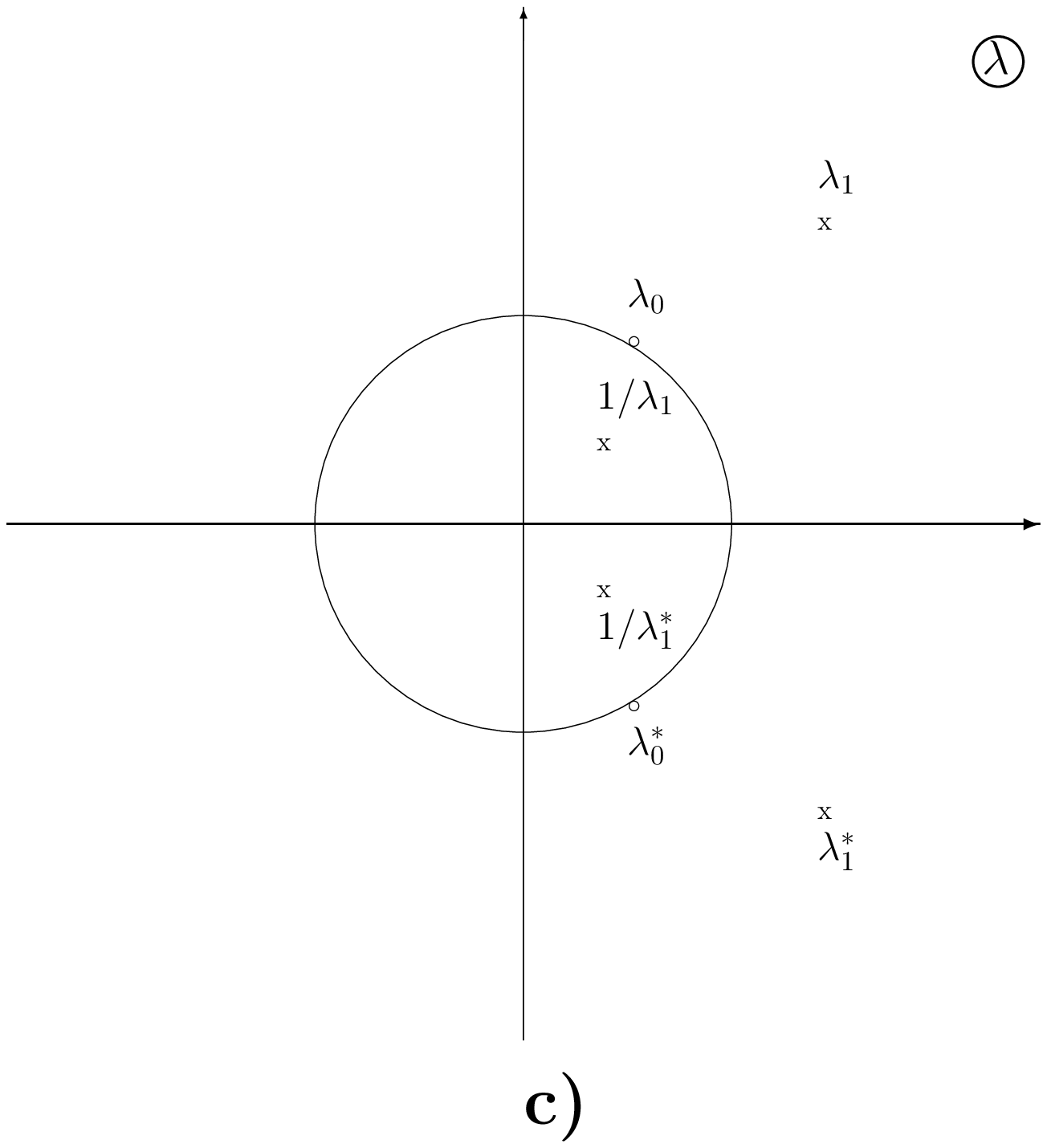}\qquad\includegraphics[width=6cm]{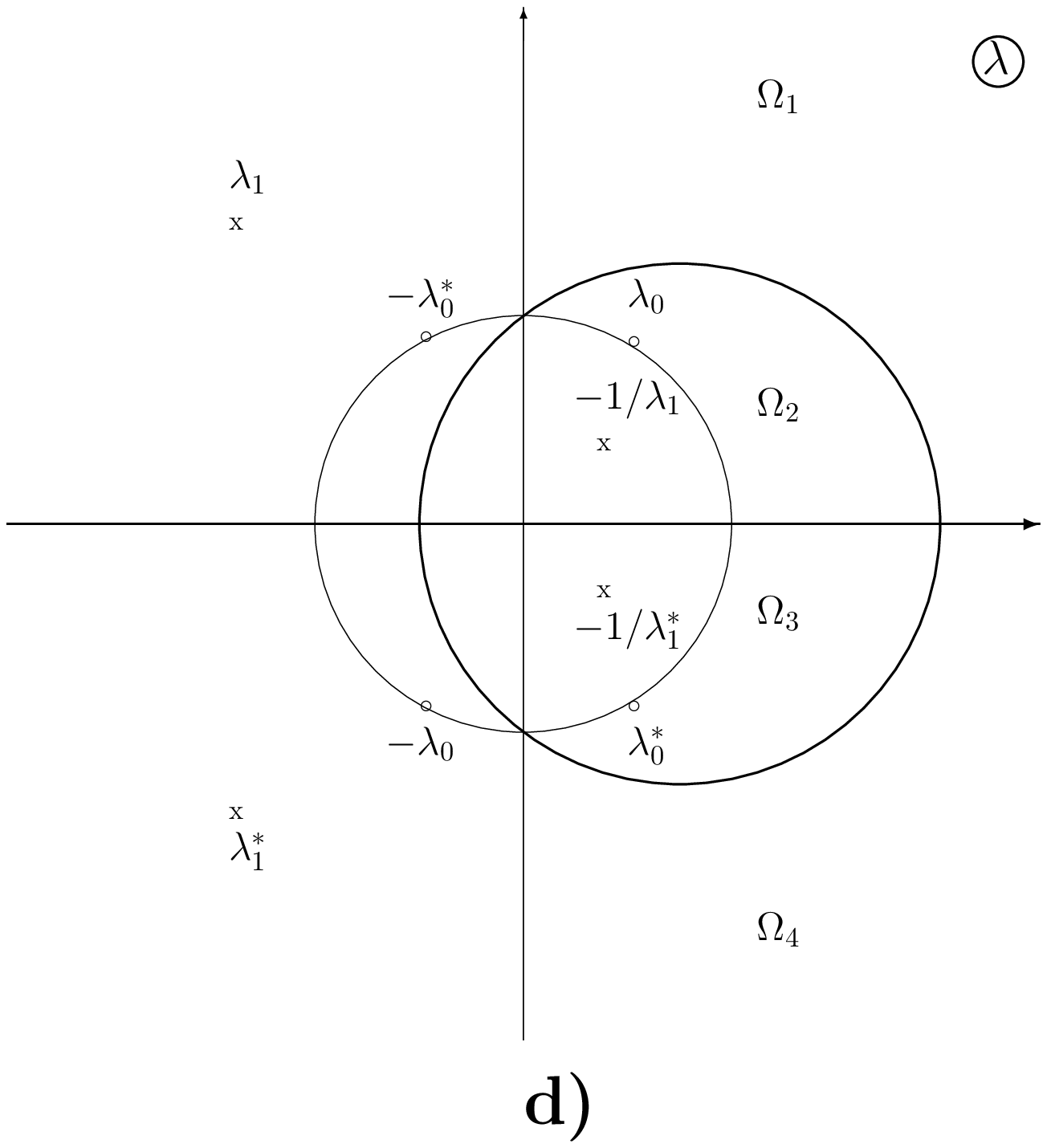}\\
  \caption{The continuous and the discrete spectrum of the operators $L_{\rm R}$ for the 4 different cases as described
  in the text. In the last case d) we have chosen $a=1/3$. }\label{fig:1}
\end{figure}

%%%%%%%%%%%%%%%
\begin{description}
  \item[Case a): $CJ \hat{C}=J$ and $\epsilon=1$.] Condition (\ref{eq:cs}) becomes:
  \begin{equation}\label{eq:Sp-a}\begin{split}
  \lambda_1( \lambda_0^2 +\lambda_1^2 -1) =0, \\
  \end{split}\end{equation}
Thus the continuous spectrum of $L_{\rm i)}$ consists of $\bbbr \cup \bbbs^1$, where $\bbbs^1$ is the unit circle
with center at the origin.

  The discrete spectrum of $L_{\rm R}$ contains quadruplets  of discrete eigenvalues. The generic quadruplet of
  eigenvalues consists of $\lambda_k$, $\lambda_k^*$,  $1/\lambda_k$ and $1/\lambda_k^*$,  see fig. \ref{fig:1}a).

  \item[Case b): $CJ \hat{C}=J$ and $\epsilon= -1$.] The analog of eq. (\ref{eq:cs}) is:
  \begin{equation}\label{eq:cs2}\begin{split}
   \im \left(  \frac{ J\xi}{\lambda -a}  + \frac{ CJC^{-1}\xi}{-\lambda^{-1} -a}  \right) =0.
  \end{split}\end{equation}
  Its solution is
  \begin{equation}\label{eq:Sp-b}\begin{split}
  \lambda_1( \lambda_0^2 +\lambda_1^2 +1) =0,
  \end{split}\end{equation}
    The second factor $\lambda_0^2 +\lambda_1^2 +1$ is always positive, therefore in this case
    the continuous spectrum of $L_{\rm ii)}$ consists of the real axis $\bbbr $ only.

  The discrete spectrum of $L_{\rm R}$ consists of quadruplets and doublet discrete eigenvalues. The generic quadruplet of
  eigenvalues consists of $\lambda_k$, $\lambda_k^*$,  $-1/\lambda_k$ and $-1/\lambda_k^*$. These quadruplets
  do not degenerate even on the unit circle. Doublet eigenvalues takes place
  only at $ i$ and $-i$, see fig. \ref{fig:1}b).

  \item[Case c): $CJ \hat{C}=-J$ and $\epsilon=1$.] From eq. (\ref{eq:cs}) we get:
  \begin{equation}\label{eq:Sp-c}\begin{split}
  \lambda_1 \left( \left( \lambda_0 -\frac{ 2a}{1+a^2}\right) ^2 +\lambda_1^2 +c_0 \right) =0, \qquad c_0 =\frac{ (1-a^2)^2}{(1+a^2)^2}.
  \end{split}\end{equation}
  Again the second factor $\lambda_0^2 +\lambda_1^2 +c_0$ is always positive, and therefore
    the continuous spectrum of $L_{\rm R}$ consists of the real axis $\bbbr $ only.

  The discrete spectrum of $L_{\rm R}$ consists of quadruplets and doublet discrete eigenvalues. The generic quadruplet
  eigenvalues consists of $\lambda_k$, $\lambda_k^*$,  $1/\lambda_k$ and $1/\lambda_k^*$. The doublet eigenvalues take place
  if $|\lambda_k|=1$, i.e. they lie on the unit circle, see fig. \ref{fig:1}c).

  \item[Case d): $CJ \hat{C}=-J$ and $\epsilon= -1$.] From eq. (\ref{eq:cs2}) we find:
  \begin{equation}\label{eq:Sp-d}\begin{split}
  \lambda_1 \left( \left( \lambda_0 -\frac{ 2a}{1-a^2}\right) ^2 +\lambda_1^2 -c_1^2 \right) =0, \qquad c_1 = \frac{ a^2+1}{|a^2-1|}.
  \end{split}\end{equation}

Thus the continuous spectrum of $L_{\rm iv)}$ consists of $\bbbr \cup \bbbs^1$, where $\bbbs^1$ is a circle
with center on the real axis at $2a/(1-a^2)$ and radius $c_1$.

  The discrete spectrum of $L_{\rm R}$ consists of quadruplets. The generic quadruplet
  eigenvalues consists of $\lambda_k$, $\lambda_k^*$,  $-1/\lambda_k$ and $-1/\lambda_k^*$. The only
  possible doublet eigenvalues at $\pm i$ are ruled out because they lie on the continuous spectrum of $L_{\rm R}$,
  see fig. \ref{fig:1}d).

\end{description}

\begin{remark}\label{rem:2}
Note that for the NJLVL models with $\mathfrak{g}\simeq su(N) $ and $J={\rm \diag} (1,0,\dots,0)$
only cases a) and b) are relevant. Indeed, there are no automorphisms of $su(N)$ that transform 
$J$ into $-J$. 
\end{remark}

In all the cases described above  one should avoid discrete eigenvalues lying on the
  continuous spectrum of $L_{\rm R}$.

Now we construct the FAS using the Gauss factors in (\ref{eq:gauR}):
\begin{equation}\label{eq:fas2}\begin{split}
\chi^\pm(x,t,\lambda) = \Psi_-(x,t,\lambda)S_J^\pm (t,\lambda)= \Psi_+(x,t,\lambda)T_J^\pm (t,\lambda) D_J^\pm (\lambda).
\end{split}\end{equation}
For the cases b) and c) $\chi^+(x,t,\lambda)$ and $\chi^-(x,t,\lambda)$ are analytic
 for $\lambda\in \bbbc_+$ and  $\lambda\in \bbbc_-$ respectively. For the cases a) and d)
 $\chi^+(x,t,\lambda)$ is analytic for  $\lambda \in \Omega_1\cup \Omega_3$ and $\chi^-(x,t,\lambda)$
-- for  $\lambda \in \Omega_2\cup \Omega_4$.

The FAS (\ref{eq:fas2}) satisfy a RHP on a contour in $\bbbc$ which coincides with the
continuous spectrum of $L_{\rm R}$:
\begin{equation}\label{eq:rhpR}\begin{split}
\chi^+(x,t,\lambda) = \chi^-(x,t,\lambda) G_J(\lambda,t), \qquad G_J(\lambda,t)= \hat{S}_J^-(\lambda,t) S_J^+(\lambda,t),
\qquad \lambda \in \mathcal{S},
\end{split}\end{equation}
where $\mathcal{S}$ is the continuous spectrum of $L_{\rm R}$, see fig. \ref{fig:1}.
This fact allows one to apply the Zakharov-Shabat dressing method for constructing the soliton
solutions of the $\bbbz_2$-reduced spinor models, very much along the ideas of \cite{zm1}.
Unfortunately now there is no natural point in $\bbbc$ at which
the RHP can be  normalized, which presents an additional difficulty in applying the dressing method.

\section{Conclusion}

We have proposed a new class of $\bbbz_2$-reduced spinor models. The spectral
properties and the construction of the FAS for their reduced Lax operators $L_{\rm R}$ are
outlined.

Other important developments are related to the interpretation of the ISM as a
generalized Fourier transform \cite{AKNS,GVY}. This can be done using the Wronskian relations to
analyze the mapping between the potential $U_{\rm R}$ and the scattering data.
The soliton solutions of these models  can be calculated using the method of \cite{zm1}
and will be published elsewhere.

New classes of generalized GN-type spinor models can be constructed choosing appropriate 
rank-2 matrices for $J$ instead of eq. (\ref{eq:J0}). Such models will have $4N$ independent
components and the inverse scattering problem for their Lax operators will be regular. 

One can also consider reductions with automorphisms $C$ such that $CJ \hat{C}\neq \pm J$.

Another important problem will be to explore the supersymmetric generalizations of the above
models.

\section*{Acknowledgements}

I am grateful to Professor A. V. Mikhailov and Professor A. S. Sorin for useful suggestions and discussions.
I also acknowledge a grant with the JINR, which allowed me to  work
on the topic 01-3-1073-2009/2013 of Dubna scientific plan and to participate in the
XV SYMPHYS conference  in Dubna.

\end{document}